# One-electron states and interband optical absorption in single-wall carbon nanotubes

**Vadim Adamyan and Sergey Tishchenko**
Department of Theoretical Physics, I.I. Mechnikov Odessa National University,
2 Dvoryanskaya Street, Odessa 65026, Ukraine

E-mail: vadamyan@onu.edu.ua , tsv@onu.edu.ua

**Abstract.** Explicit expressions for the wave functions and dispersion equation for the band $\pi$ - electrons in single-wall carbon nanotubes are obtained within the method of zero-range potentials. They are then used to investigate the absorption spectrum of polarized light caused by direct interband transitions in isolated nanotubes. It is shown that, at least, under the above approximations, circular dichroism is absent in chiral nanotubes for a light wave propagating along the tube axis. The results obtained are compared with those calculated in a similar way for a graphite plane.

## 1. Introduction

The present work is devoted to interpretation of the optical spectra of isolated carbon nanotubes (CNTs). The problem of identification of nanotubes by their spectra is very important in view of a large scale of their applications. As an example of investigation of the optical absorption spectra for direct interband transitions in CNTs, we develop here a general approach to description of the electronic structure of nanotubes, which may be useful for identification of certain nanotubes by their characteristic contributions to observable optical spectra. As a rather simple and convenient tool for obtaining the related explicit analytic expressions and numerical calculations, we propose here the method of zero-range potentials (ZRPs) [1,2] that has already demonstrated its efficacy in investigations of the band structure of nonchiral nanotubes [3]. We use it as a suitable approximation for description of the one-particle states of CNT $\pi$-electrons. In doing so, we describe the periodic structure of the nanotubes by a small number of identical monoatomic spirals shifted with respect to each other. Note that the ZRP method in application to the study of nanotubes allows one, in particular, to take into account their spatial structure in full without any restrictions on the number of interacting neighbours. Such an approach essentially simplifies calculations of the electron structure of chiral nanotubes, which may contain tens and hundreds of atoms in a unit cell. At the same time, in many cases only two spirals are enough to model the spatial structure of nanotubes. The representation of a tube as an aggregate of monoatomic spirals leads to an extended zone scheme, with the number of bands equalling the minimal number of spirals necessary for complete description of the nanotube structure, contrary to the standard zone scheme, in which the number of bands equals the number of atoms in the unit cell. However, for nonchiral nanotubes, due to their special symmetry, such an approach offers no advantages in comparison with the standard description based on a direct account of the translation symmetry of tubes [3].

In the next section of this paper, we describe the spiral structure of chiral tubes. In section 3, we deduce the dispersion equations for the bands of $\pi$-electrons and find explicit expressions for the corresponding wave functions by using the ZRP method and the obtained parametrization of spiral structure of tubes. The forbidden band widths for various nanotubes that we obtain in this way prove to be the same, within an experimental accuracy, as those obtained in [4] by means of scanning tunnelling spectroscopy. The results of section 3 are then used in section 4 to derive a general expression for the contribution to the light absorption coefficient due to direct interband transitions in isolated nanotubes. In subsections 4.1-4.3, expressions for the matrix elements of the operator of electron - photon interaction for direct interband transitions are obtained and, as a result, the selection rules for different polarizations of the incident light are derived. In subsection 4.4, the general formulae are used to calculate the absorption coefficient for nanotube (15,14). The fact that the diameter of the latter coincides with the average diameter of the tubes experimentally obtained and investigated in [5] allowed us to conduct comparison of our theoretical absorption lines and those obtained in [5]. To explain the similarities and discrepancies between our theory and experiment, and to further check out the robustness of the developed approach, the same method is used in the section 5 to investigate the optical absorption spectrum for the infinite graphite plane.

## 2. Spiral structure of carbon nanotubes

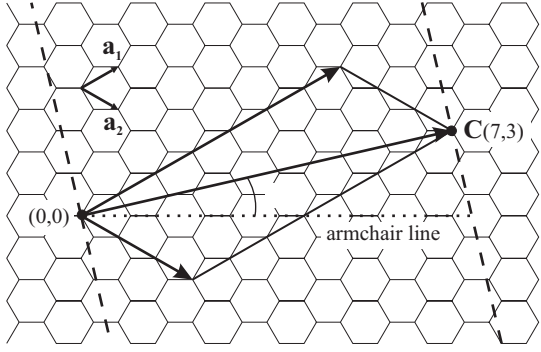

**Figure 1.** Nanotubes can be obtained by cutting strips out of a monoatomic graphite layer along different parallel dashed lines and then wrapping them to form cylinders.

A single-wall carbon nanotube can be represented as a strip of graphite plane wrapped in the form of a cylinder. The vector $\mathbf{C}(n,m) = n\mathbf{a}_1 + m\mathbf{a}_2$, where $\mathbf{a}_1$ and $\mathbf{a}_2$ are the unit vectors of the hexagonal lattice ($|\mathbf{a}_1| = |\mathbf{a}_2| = \sqrt{3}b$, $b = 0.142$ nm is the bond length) and $n$ and $m$ are the integers uniquely identifying a nanotube, is perpendicular to the nanotube axis and joins two points $(0,0)$ and $(n, m)$, which coincide as the strip is wrapped into a cylinder. The numbers $(n, m)$ determine the nanotube radius and also the chiral angle, that is, the angle between the vectors $\mathbf{C}$ and $\mathbf{a}_1 + \mathbf{a}_2$ (armchair line), shown in figure 1, as follows

$$R = \frac{\sqrt{3}b\sqrt{n^2 + nm + m^2}}{2\pi}, \qquad \cos\theta = \frac{\sqrt{3}}{2}\frac{n+m}{\sqrt{n^2+nm+m^2}}.$$

As a nanotube is a long cylindrical molecule whose length is 1000 times greater than its diameter, it is natural to consider it as a quasionedimensional crystal with the unit cell in the form of a ring of a certain width. It is possible to reproduce all the nanotube by translating this cell along the tube axis. However, chiral nanotubes may contain tens and even hundreds of atoms in such a cell. It is more convenient to consider such nanotubes as an aggregate of a small number of displaced monoatomic spirals. The order of the rotational axis of the nanotube is equal to the greatest common divisor of the nanotube indices $n$ and $m$, $M = \mathrm{GCD}(n,m)$ [6]. The structure of such a tube can be described by rotating the pair of spirals through angles that are multiples of $\alpha = 2\pi/M$. That is, to describe the structure of an arbitrary nanotube, $M$ pairs of spirals are necessary. Thus, implying the axis $z$ to be parallel to the tube axis, the positions of the atoms can be described in the Cartesian system by their radius vectors

$$\mathbf{r}_{st}^l = (R\cos(s\varphi + t\alpha + \delta_l), R\sin(s\varphi + t\alpha + \delta_l), sd + b_l)$$

$$s = 0, \pm 1, \pm 2, \ldots, \pm\infty, \qquad t = 0, 1, \ldots, M-1, \qquad l = 0, 1.$$

Here, $s$ numbers the atoms in the spiral, $t$ numbers the pair of spirals, $l$ distinguishes spirals within one pair, $\delta_l$ and $b_l$ are respectively the rotational displacement about and the shift along the nanotube axis for each spiral in the pair: $\delta_0 = 0$, $\delta_1 = R^{-1}a(\cos\theta_1 + \cos\theta_2)2/3$, $b_0 = 0$, $b_1 = a(\sin\theta_1 - \sin\theta_2)2/3$, where $\theta_1$ and $\theta_2$ are the angles between the vectors $\mathbf{a}_1$ and $\mathbf{C}$ and $\mathbf{a}_2$ and $\mathbf{C}$, respectively. The latter are determined by expressions

$$\cos\theta_1 = \frac{n + m/2}{\sqrt{n^2+nm+m^2}}, \quad \sin\theta_1 = \frac{\sqrt{3}}{2}\frac{m}{\sqrt{n^2+nm+m^2}}; \quad \cos\theta_2 = \frac{m + n/2}{\sqrt{n^2+nm+m^2}}, \quad \sin\theta_2 = \frac{\sqrt{3}}{2}\frac{n}{\sqrt{n^2+nm+m^2}}.$$

The transition from one atom to its next neighbour inside one spiral is attained by rotating through the angle $\varphi = R^{-1}a(n_t\cos\theta_1 + m_t\cos\theta_2)$ and by shifting along the axis by $d = a(n_t\sin\theta_1 - m_t\sin\theta_2)$. $n_t$ and $m_t$ are the integers determined by the indices $n$ and $m$:

$$n_t = -\frac{2m+n}{\Delta\widetilde{R}} + \frac{r}{\Delta}n, \qquad m_t = \frac{2n+m}{\Delta\widetilde{R}} + \frac{r}{\Delta}m,$$

where $\widetilde{R} = 3$ if $(n-m)/3M$ is an integer and $\widetilde{R} = 1$ otherwise, $\Delta = 2\dfrac{n^2 + nm + m^2}{M\widetilde{R}}$,

$r = \dfrac{\Delta}{M}\mathrm{Fr}\left[\dfrac{M}{\Delta\widetilde{R}}\left(3 - 2\dfrac{n-m}{n}\right) + \dfrac{M}{n}\left(\dfrac{n-m}{M}\right)^{\varphi(n/M)-1}\right]$. Here $\mathrm{Fr}[x] = x - [x]$ is the fractional part of the rational number $x$, and $\varphi(a)$ is the Euler function, giving the number of coprimes less than $a$ [6].

## 3. Spectrum and wave functions of band electrons

Recall that carbon atoms have four valence electrons. In a tube, three of them form σ-bonds and the last one forms a delocalized π-bond. The electronic properties of nanotubes are determined mainly by the π-electrons. We model states of the delocalized π-electrons in a tube by using the method of zero-range potentials (ZRP) [1,2]. The essence of this method is that the interaction of an electron with the atoms or ions of a lattice is described not by a potential but by boundary conditions imposed on the one-electron wave function at a discrete set of points of the atoms location:

$$\lim_{\rho_{st}^l \to 0}\left[\dfrac{\partial}{\partial \rho_{st}^l}(\rho_{st}^l \Psi) - \lambda \rho_{st}^l \Psi\right] = 0, \quad \rho_{st}^l = |\mathbf{r} - \mathbf{r}_{st}^l|. \quad (1)$$

Recall that boundary condition (1) is, in a sense, the three-dimensional analog of the δ-interaction or an equivalent expression for the Enrico Fermi pseudo-potential $-\lambda^{-1}\delta(\rho)\dfrac{\partial}{\partial \rho}\rho\bigg|_{\rho=0+}$ [2]. The parameter $\lambda = -\sqrt{\dfrac{2\mu}{\hbar^2}|E_0|}$ in (1) is the interaction constant, where $\mu$ is the universal fitting parameter with the dimension of mass (the same for all carbon nanotubes!), and $E_0$ is the ionization energy of an isolated carbon atom. The parameter $\mu$ is chosen from the requirement that the distances between two nearest Van Hove singularities, which are on different sides from the edges of a narrow forbidden band in the zigzag nanotube (15,0), obtained theoretically within the framework of the considered model [3] and measured experimentally in [4] coincide.

We assume that the wave function satisfies, in addition to (1), the Schrödinger equation for a free particle

$$-\dfrac{\hbar^2}{2\mu}\Delta\Psi = E\Psi.$$

Let us seek the wave function in the form

$$\Psi(\mathbf{r}, E) = \sum_{l=0}^{1}\sum_{t=0}^{M-1}\sum_{s=-\infty}^{+\infty} C_{st}^l G(\gamma, |\mathbf{r} - \mathbf{r}_{st}^l|), \quad \gamma = \sqrt{-\dfrac{2\mu}{\hbar^2}E}, \quad (2)$$

where $G(\gamma, x) = \dfrac{e^{-\gamma x}}{x}$.

Substituting this expression in (1), we obtain for coefficients $C_{st}^l$ a homogeneous system

$$(-\gamma - \lambda)C_{s't'}^{l'} + \sum_{\substack{s=-\infty \\ s \neq s'}}^{+\infty} C_{st'}^{l'} G(\gamma, |\mathbf{r}_{s't'}^{l'} - \mathbf{r}_{st'}^{l'}|) + \sum_{\substack{t=0 \\ t \neq t'}}^{M-1}\sum_{s=-\infty}^{+\infty} C_{st}^{l'} G(\gamma, |\mathbf{r}_{s't'}^{l'} - \mathbf{r}_{st}^{l'}|) + \sum_{\substack{l=0 \\ l \neq l'}}^{1}\sum_{t=0}^{M-1}\sum_{s=-\infty}^{+\infty} C_{st}^l G(\gamma, |\mathbf{r}_{s't'}^{l'} - \mathbf{r}_{st}^l|) = 0, \quad (3)$$

$$s' = 0, \pm 1, \pm 2, ..., \pm\infty, \quad t' = 0, 1, ..., M-1, \quad l' = 0, 1.$$

According to the symmetry of the considered system, we represent these coefficients as $C_{st}^l = C_0^l e^{-ips} e^{-i\alpha v t}$, where $\alpha = 2\pi/M$, $\nu = 0, 1, ..., M-1$, $p \in [-\pi, \pi)$. This yields a system of two equations

$$H(\gamma, p, \nu)C_0^{l'} + \sum_{\substack{l=0 \\ l \neq l'}}^{1} C_0^l E_{l'l}(\gamma, p, \nu) = 0, \qquad l' = 0, 1, \quad (4)$$

where

$$H(\gamma,p,\nu) = -\gamma - \lambda + \sum_{\substack{s=-\infty \\ s\neq 0}}^{+\infty} G\left(\gamma,\sqrt{2R^2[1-\cos(s\varphi)]+(sd)^2}\right)e^{ips}$$

$$+ (1-\delta_{M,1})\sum_{t=1}^{M-1}\sum_{s=-\infty}^{+\infty} G\left(\gamma,\sqrt{2R^2[1-\cos(s\varphi+t\alpha)]+(sd)^2}\right)e^{ips}e^{i\alpha\nu t} \quad,$$

$$E_{l'l}(\gamma,p,\nu) = \sum_{t=0}^{M-1}\sum_{s=-\infty}^{+\infty} G\left(\gamma,\sqrt{2R^2[1-\cos(s\varphi+t\alpha+\delta_{l'}-\delta_l)]+(sd+b_{l'}-b_l)^2}\right)e^{ips}e^{i\alpha\nu t}.$$

The solvability condition for (4) gives for negative energies the following dispersion equations for π-electrons band states in the nanotube:

$$H(\gamma,p,\nu) \pm |E_{01}(\gamma,p,\nu)| = 0 \quad, \quad (E_{01} = \overline{E_{10}}). \tag{5}$$

Thus $C_0^l = C_0 F_l(\gamma,p,\nu)$, where $F_l(\gamma,p,\nu)$ is determined from (4) ($F_0(\gamma,p,\nu) = 1$, $F_1(\gamma,p,\nu) = -H(\gamma,p,\nu)/E_{01}(\gamma,p,\nu)$) and $C_0(\gamma,p,\nu)$ is a normalization constant assuring the normalization of Bloch functions (2) to the δ-function. As

$$\int \overline{\Psi(\mathbf{r},E(p,\nu))}\Psi(\mathbf{r},E(p',\nu'))d\mathbf{r} = \sum_{l,l'=0}^{1}\sum_{t,t'=0}^{M-1}\sum_{s,s'=-\infty}^{+\infty} \overline{C_{st}^l(\gamma,p,\nu)}C_{s't'}^{l'}(\gamma,p',\nu')\int G(\gamma,|\mathbf{r}-\mathbf{r}_{st}^l|)G(\gamma,|\mathbf{r}-\mathbf{r}_{s't'}^{l'}|)d\mathbf{r}$$

and

$$\int G(\gamma,|\mathbf{r}-\mathbf{r}_{st}^l|)G(\gamma,|\mathbf{r}-\mathbf{r}_{s't'}^{l'}|)d\mathbf{r} = \frac{2\pi}{\gamma}\exp(-\gamma|\mathbf{r}_{st}^l - \mathbf{r}_{s't'}^{l'}|),$$

then

$$\int \overline{\Psi(\mathbf{r},E(p,\nu))}\Psi(\mathbf{r},E(p',\nu'))d\mathbf{r} = |C_0|^2 \frac{2\pi}{\gamma}\sum_{l,l'=0}^{1}\overline{F_l(\gamma,p,\nu)}F_{l'}(\gamma,p,\nu)N_{ll'}(\gamma,p,\nu)\delta(p-p')\delta_{\nu,\nu'},$$

where

$$N_{ll'}(\gamma,p,\nu) = \sum_{t=0}^{M-1}\sum_{s=-\infty}^{+\infty}\exp\left(-\gamma\sqrt{2R^2[1-\cos(s\varphi+t\alpha+\delta_l-\delta_{l'})]+(sd+b_l-b_{l'})^2}\right)e^{ips}e^{i\alpha\nu t}.$$

Therefore 
$$C_0(\gamma,p,\nu) = \left(\frac{2\pi}{\gamma}\sum_{l,l'=0}^{1}\overline{F_l(\gamma,p,\nu)}F_{l'}(\gamma,p,\nu)N_{ll'}(\gamma,p,\nu)\right)^{-\frac{1}{2}}.$$

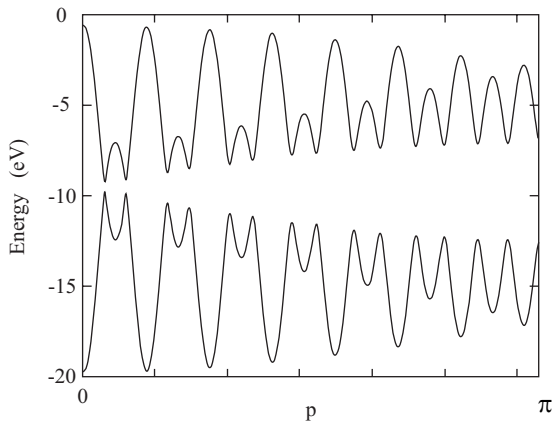

**Figure 2.** The zone scheme of nanotube (15,14) in natural spiral coordinates, that is, the extended zone scheme.

As an example, we consider a nanotube (15,14) with diameter $d = 1.966$ nm and chiral angle $\theta = 1.141°$. It is sufficient for two spirals (M=1) to describe its spatial structure; accordingly, two bands are obtained. At the same time, such a nanotube contains 2524 atoms per unit cell and hence the same number of bands in the standard zone scheme, some of the bands clamping. Evidently, analysing two bands by using the extended zone scheme (figure 2) is much easier than analysing 2524 bands. In figure 2, the band gap ($E_g = 0.42\ eV$) and the extremum points that generate the Van Hove singularities in the electronic density of states inherent to quasionedimensional systems are explicitly visible.

Table 1. Comparison of experimental results [4] with results obtained by the ZRP method.

| | Experiment | | | | ZRP | | | |
|---|---|---|---|---|---|---|---|---|
| № | diameter ±0,1(nm) | angle ±1(deg) | $E_g$ (eV) | Type | (n,m) | diameter (nm) | Angle (deg) | $E_g$ (eV) |
| 1 | 1.4 | 25 | 0.55 | semiconductor | (17,2) | 1.416 | 24.504 | metal |
| | | | | | (18,2) | 1.494 | 24.791 | 0.551 |
| | | | | | (16,2) | 1.338 | 24.182 | 0.62 |
| 2 | 1.4 | 4 | 0.6 | semiconductor | (11,9) | 1.358 | 3.304 | 0.608 |
| 16 | 1.4 | 4 | 0.5 | semiconductor | (12,9) | 1.429 | 4.715 | metal |
| | | | | | (12,10) | 1.494 | 3.004 | 0.553 |
| 3 | 2 | 7 | 0.5 | semiconductor | (18,12) | 2.048 | 6.587 | metal |
| | | | | | (18,11) | 1.985 | 7.934 | 0.416 |
| | | | | | (17,11) | 1.913 | 7.053 | 0.418 |
| 4 | 1.2 | 24 | 0.65 | semiconductor | (14,2) | 1.182 | 23.413 | metal |
| | | | | | (15,2) | 1.26 | 23.822 | 0.652 |
| 5 | 1.7 | 9 | 1.7 | metal | (16,9) | 1.717 | 9.183 | semiconductor |
| | | | | | (15,9) | 1.644 | 8.213 | 1.453 |
| 6 | 1.3 | 14 | 1.8 | metal | (13,5) | 1.26 | 14.392 | semiconductor |
| | | | | | (14,5) | 1.336 | 15.295 | 1.74 |
| | | | | | (14,6) | 1.392 | 13.004 | semiconductor |
| 7 | 1.1 | 30 | 1.9 | metal | (15,0) | 1.174 | 30 | 1.9 |
| | | | | | (14,0) | 1.096 | 30 | semiconductor |
| | | | | | (13,0) | 1.018 | 30 | semiconductor |
| 13 | 1.5 | 7 | 2 | metal | (13,8) | 1.437 | 7.827 | semiconductor |
| | | | | | (14,9) | 1.572 | 7.154 | semiconductor |
| 14 | 1.4 | 16 | 1.9 | metal | (15,5) | 1.411 | 16.102 | semiconductor |
| | | | | | (14,5) | 1.336 | 15.295 | 1.74 |
| | | | | | (16,5) | 1.487 | 16.826 | semiconductor |
| 17 | 1.4 | 9 | 0.55 | semiconductor | (13,7) | 1.376 | 9.826 | metal |
| | | | | | (12,7) | 1.303 | 8.639 | 0.635 |
| 18 | 1.2 | 16 | 0.6 | semiconductor | (13,4) | 1.205 | 16.996 | metal |
| | | | | | (12,4) | 1.129 | 16.102 | 0.734 |
| 19 | 1.3 | 6 | 0.6 | semiconductor | (11,8) | 1.294 | 5.209 | metal |
| | | | | | (12,8) | 1.365 | 6.587 | 0.604 |
| 20 | 1.3 | 29 | 0.5 | semiconductor | (16,0) | 1.253 | 30 | 0.63 |
| | | | | | (17,0) | 1.331 | 30 | 0.643 |
| 21 | 1.9 | 16 | 0.4 | semiconductor | (20,7) | 1.9 | 15.535 | 0.434 |
| | | | | | (21,7) | 1.976 | 16.102 | 0.419 |

By means of scanning tunnelling spectroscopy it is possible to obtain the electronic density of states near the Fermi level of a nanotube and thus to find the band gap for semiconducting nanotubes or the distance between two nearest Van Hove singularities, which are on different sides from the edges of a narrow forbidden band, for "metal" nanotubes, the nanotubes' diameters and chiral angles [4]. We have compared the experimental data with data obtained by the ZRP method (Table 1). It turns out within the accuracy of experiment that several values of the nanotube indices correspond to each pair of the diameter–angle measured in [4]. Moreover, some of the appropriate indices for semiconducting tubes ($E_g \in 0.5 - 0.6 eV$) are sometimes metallic ones and vice versa, among the corresponding indices for metal tubes there are those for semiconducting tubes. One can see from the table 1 that the results of carried out calculations are in satisfactory agreement with the energy gaps data obtained in [4] by the method of scanning tunnelling spectroscopy within the accuracy of latter (0.05 ~0.1 eV).

**4. Optical absorption caused by direct interband transitions in isolated nanotubes**

The energy absorbed by a nanotube per unit of time in the field of a plane electromagnetic wave
$$\mathbf{A}(\mathbf{r},t) = \mathbf{e}A_0 \exp[i(\mathbf{kr} - \omega t)],$$
where $\mathbf{e}$ is a unit polarization vector, due to direct interband transitions from a state 1 in the valence band to a state 2 in the conduction band is, according to the Fermi golden rule,

$$E(\mathbf{e}) = \frac{4\pi\omega^2}{\hbar c^2}|A_0|^2 \sum_{\nu_1,\nu_2} \sum_{p_1,p_2} |(\mathbf{e} \cdot \mathbf{g}(p_1,\nu_1,p_2,\nu_2))|^2 \delta(E_2(p_2,\nu_2) - E_1(p_1,\nu_1) - \hbar\omega), \quad (6)$$

where $\mathbf{g}(p_1,\nu_1,p_2,\nu_2) = \left(\frac{\hbar}{2\omega}\right)^{\frac{1}{2}} \frac{ie\hbar}{m_e} \langle \Psi_1 | \nabla | \Psi_2 \rangle$ is the matrix element of electron - photon interaction [7]. Let's consider temperatures $k_B T \ll E_g$ provided that all states below the Fermi level are occupied and all states above it are empty. The ratio of energy (6) to the energy flux of the electromagnetic wave $\frac{1}{2\pi}\frac{\omega^2}{c} n |A_0|^2$, $n$ being the refraction coefficient of a medium surrounding the tube,

$$\alpha(\mathbf{e}) = \frac{8\pi^2}{c\hbar n} \sum_{\nu_1,\nu_2} \sum_{p_1,p_2} |(\mathbf{e} \cdot \mathbf{g}(p_1,\nu_1,p_2,\nu_2))|^2 \delta(E_2(p_2,\nu_2) - E_1(p_1,\nu_1) - \hbar\omega) \quad (7)$$

characterises the absorbing properties of the tube depending on the frequency and polarization of the incident radiation. From now on, we shall call this quantity the absorption coefficient of the tube.

*4.1. Parallel polarization*

Hereafter we assume that the tube axis is parallel to the z-axis. Let the incident light be polarized parallel to the nanotube axis (parallel polarization or copolarization). Using the wave functions (2) and the dispersion equation (5), we obtain the following expression for the z-component of the matrix element of electron - photon interaction

$$g_z = \left(\frac{\hbar}{2\omega}\right)^{\frac{1}{2}} \frac{ie\hbar}{m_e} \int \overline{\Psi_1} \frac{\partial}{\partial z} \Psi_2 d\mathbf{r} = \left(\frac{\hbar}{2\omega}\right)^{\frac{1}{2}} \frac{ie\hbar}{m_e} \overline{C_0(\gamma_1,p_1,\nu_1)} C_0(\gamma_2,p_2,\nu_2) \frac{-4\pi}{\gamma_2^2 - \gamma_1^2}$$
$$\times \sum_{l=0}^{1}\left[\overline{F_l(\gamma_1,p_1,\nu_1)}F_l(\gamma_2,p_2,\nu_2)Q_{00} + \sum_{\substack{l'=0 \\ l' \neq l}}^{1} \overline{F_l(\gamma_1,p_1,\nu_1)}F_{l'}(\gamma_2,p_2,\nu_2)Q_{ll'}\right]\delta_{\nu_1,\nu_2}\delta_{p_1,p_2}, \quad (8)$$

where

$$Q_{00}(\gamma_1,p_1,\nu_1,\gamma_2,p_2,\nu_2) = \sum_{\substack{s=-\infty \\ s \neq 0}}^{+\infty} sd\left[\tilde{G}\left(\gamma_1,\sqrt{2R^2[1-\cos(s\varphi)]+(sd)^2}\right) - \tilde{G}\left(\gamma_2,\sqrt{2R^2[1-\cos(s\varphi)]+(sd)^2}\right)\right]e^{ip_2 s}$$

$$+ (1-\delta_{M,1})\sum_{t=1}^{M-1}\sum_{s=-\infty}^{+\infty} sd\left[\tilde{G}\left(\gamma_1,\sqrt{2R^2[1-\cos(s\varphi+t\alpha)]+(sd)^2}\right) - \tilde{G}\left(\gamma_2,\sqrt{2R^2[1-\cos(s\varphi+t\alpha)]+(sd)^2}\right)\right]e^{ip_2 s}e^{i\alpha\nu_2 t},$$

$$Q_{ll'}(\gamma_1, p_1, \nu_1, \gamma_2, p_2, \nu_2) = \sum_{t=0}^{M-1}\sum_{s=-\infty}^{+\infty}(sd+b_l-b_{l'})\left[\widetilde{G}\left(\gamma_1, \sqrt{2R^2[1-\cos(s\varphi+t\alpha+\delta_l-\delta_{l'})]+(sd+b_l-b_{l'})^2}\right)\right.$$
$$\left.-\widetilde{G}\left(\gamma_2, \sqrt{2R^2[1-\cos(s\varphi+t\alpha+\delta_l-\delta_{l'})]+(sd+b_l-b_{l'})^2}\right)\right]e^{ip_2 s}e^{i\alpha\nu_2 t}$$

and the function $\widetilde{G}(\gamma, x) = \dfrac{e^{-\gamma x}}{x^2}\left(\gamma + \dfrac{1}{x}\right)$ is introduced.

The Kronecker deltas $\delta_{\nu_1,\nu_2}\delta_{p_1,p_2}$ in (8) determine the selection rules $\nu_1 = \nu_2$, $p_1 = p_2$. Representing the matrix element as $g_z(p_1,\nu_1,p_2,\nu_2) = \widetilde{g}_z(p_1,\nu_1)\delta_{\nu_1,\nu_2}\delta_{p_1,p_2}$ and replacing the sum in $p$ in (7) by an integral, we obtain, with account of the spin degeneracy, that the tube absorption coefficient for the parallel polarization is

$$\alpha_{\|} = \frac{8\pi}{c\hbar n}\frac{L}{T}\sum_{\nu=0}^{M-1}\int|\widetilde{g}_z(p,\nu)|^2 \delta(E_2(p,\nu) - E_1(p,\nu) - \hbar\omega)\,dp,$$

where L is the nanotube length, $T = b\sqrt{\dfrac{9\Delta}{2M\widetilde{R}}}$ is the length of the elementary cell of the nanotube. And, finally,

$$\alpha_{\|}(\omega) = \frac{8\pi}{c\hbar n}\frac{L}{T}\sum_{\nu=0}^{M-1}\frac{|\widetilde{g}_z(p,\nu)|^2}{\left|\dfrac{\partial}{\partial p}(E_2(p,\nu) - E_1(p,\nu))\right|}\Bigg|_{p=p(\omega,\nu)},$$

where $p(\omega,\nu)$ is found from the condition $E_2(p,\nu) - E_1(p,\nu) = \hbar\omega$.

### 4.2. Circular right polarization

Let now the light be propagating along the nanotube axis and have the circular right polarization. Then the corresponding matrix element is

$$g^R = \left(\frac{\hbar}{2\omega}\right)^{\frac{1}{2}}\frac{ie\hbar}{m_e}\int \Psi_1\left(\frac{\partial}{\partial x}+i\frac{\partial}{\partial y}\right)\Psi_2 d\mathbf{r} = \left(\frac{\hbar}{2\omega}\right)^{\frac{1}{2}}\frac{ie\hbar}{m_e}\overline{C_0(\gamma_1,p_1,\nu_1)}C_0(\gamma_2,p_2,\nu_2)\frac{-4\pi R}{\gamma_2^2-\gamma_1^2}$$
$$\times \sum_{l=0}^{1}\left[\overline{F_l(\gamma_1,p_1,\nu_1)}F_l(\gamma_2,p_2,\nu_2)Q_{ll}+\sum_{\substack{l'=0\\l'\neq l}}^{1}\overline{F_l(\gamma_1,p_1,\nu_1)}F_{l'}(\gamma_2,p_2,\nu_2)Q_{ll'}\right]\delta_{\nu_1+1,\nu_2}\delta_{p_1+\varphi,p_2},$$

where
$$Q_{ll}(\gamma_1,p_1,\nu_1,\gamma_2,p_2,\nu_2) =$$
$$= \sum_{\substack{s=-\infty\\s\neq 0}}^{+\infty}(1-e^{-is\varphi})e^{i\delta_l}\left[\widetilde{G}\left(\gamma_1,\sqrt{2R^2[1-\cos(s\varphi)]+(sd)^2}\right)-\widetilde{G}\left(\gamma_2,\sqrt{2R^2[1-\cos(s\varphi)]+(sd)^2}\right)\right]e^{ip_2 s}$$
$$+(1-\delta_{M,1})\sum_{t=1}^{M-1}\sum_{s=-\infty}^{+\infty}(1-e^{-i(s\varphi+\alpha t)})e^{i\delta_l}\left[\widetilde{G}\left(\gamma_1,\sqrt{2R^2[1-\cos(s\varphi+t\alpha)]+(sd)^2}\right)\right.$$
$$\left.-\widetilde{G}\left(\gamma_2,\sqrt{2R^2[1-\cos(s\varphi+t\alpha)]+(sd)^2}\right)\right]e^{ip_2 s}e^{i\alpha\nu_2 t},$$

$$Q_{ll'}(\gamma_1,p_1,\nu_1,\gamma_2,p_2,\nu_2) = \sum_{t=0}^{M-1}\sum_{s=-\infty}^{+\infty}(1-e^{-i(s\varphi+t\alpha+\delta_l-\delta_{l'})})e^{i\delta_l}\left[\widetilde{G}\left(\gamma_1,\sqrt{2R^2[1-\cos(s\varphi+t\alpha+\delta_l-\delta_{l'})]+(sd+b_l-b_{l'})^2}\right)\right.$$
$$\left.-\widetilde{G}\left(\gamma_2,\sqrt{2R^2[1-\cos(s\varphi+t\alpha+\delta_l-\delta_{l'})]+(sd+b_l-b_{l'})^2}\right)\right]e^{ip_2 s}e^{i\alpha\nu_2 t}.$$

In this case, the selection rules are $\nu_2 = \nu_1 + 1$, $p_2 = p_1 + \varphi$.
Substituting

$$g^R(p_1,\nu_1,p_2,\nu_2) = \tilde{g}^R(p_1,\nu_1,p_1+\varphi,\nu_1+1)\tilde{\delta}_{\nu_1+1,\nu_2}\delta_{p_1+\varphi,p_2}, \quad \tilde{\delta}_{\nu_1+1,\nu_2} = \begin{cases} 1, & \nu_2 = \nu_1+1 \pmod{M} \\ 0, & \nu_2 \neq \nu_1+1 \pmod{M} \end{cases}$$

into (7) we get the following expression for the absorption coefficient $\alpha^R$ of the circular right polarized light propagating along the tube axis:

$$\alpha^R(\omega) = \frac{8\pi}{c\hbar n}\frac{L}{T}\sum_{\nu=0}^{M-1}\frac{|\tilde{g}^R(p,\nu,p+\varphi,\nu+1)|^2}{\left|\frac{\partial}{\partial p}(E_2(p+\varphi,\nu+1) - E_1(p,\nu))\right|}\Bigg|_{p=p(\omega,\nu)},$$

where $p(\omega,\nu)$ is found from the condition $E_2(p+\varphi,\nu+1) - E_1(p,\nu) = \hbar\omega$.

### 4.3. Circular left polarization

If the light is propagating along the nanotube axis and has circular left polarization, then the corresponding matrix element is

$$g^L = \left(\frac{\hbar}{2\omega}\right)^{1/2}\frac{ie\hbar}{m_e}\int\Psi_1\left(\frac{\partial}{\partial x} - i\frac{\partial}{\partial y}\right)\Psi_2 d\mathbf{r} = \left(\frac{\hbar}{2\omega}\right)^{1/2}\frac{ie\hbar}{m_e}\overline{C_0(\gamma_1,p_1,\nu_1)}C_0(\gamma_2,p_2,\nu_2)\frac{-4\pi R}{\gamma_2^2 - \gamma_1^2}$$

$$\times \sum_{l=0}^{M-1}\left[\overline{F_l(\gamma_1,p_1,\nu_1)}F_l(\gamma_2,p_2,\nu_2)Q_{ll} + \sum_{\substack{l'=0 \\ l'\neq l}}^{M-1}\overline{F_l(\gamma_1,p_1,\nu_1)}F_{l'}(\gamma_2,p_2,\nu_2)Q_{ll'}\right]\delta_{\nu_1-1,\nu_2}\delta_{p_1-\varphi,p_2},$$

$$Q_{ll}(\gamma_1,p_1,\nu_1,\gamma_2,p_2,\nu_2) =$$
$$= \sum_{\substack{s=-\infty \\ s\neq 0}}^{+\infty}(1-e^{is\varphi})e^{-i\delta_l}\left[\tilde{G}\left(\gamma_1,\sqrt{2R^2[1-\cos(s\varphi)] + (sd)^2}\right) - \tilde{G}\left(\gamma_2,\sqrt{2R^2[1-\cos(s\varphi)] + (sd)^2}\right)\right]e^{ip_2 s} +$$

$$+ (1-\delta_{M,1})\sum_{t=1}^{M-1}\sum_{s=-\infty}^{+\infty}(1-e^{i(s\varphi+t\alpha)})e^{-i\delta_l}\left[\tilde{G}\left(\gamma_1,\sqrt{2R^2[1-\cos(s\varphi+t\alpha)] + (sd)^2}\right)\right.$$
$$\left. - \tilde{G}\left(\gamma_2,\sqrt{2R^2[1-\cos(s\varphi+t\alpha)] + (sd)^2}\right)\right]e^{ip_2 s}e^{i\alpha\nu_2 t},$$

$$Q_{ll'}(\gamma_1,p_1,\nu_1,\gamma_2,p_2,\nu_2) = \sum_{s=-\infty}^{+\infty}(1-e^{i(s\varphi+t\alpha+\delta_l-\delta_{l'})})e^{-i\delta_l}\left[\tilde{G}\left(\gamma_1,\sqrt{2R^2[1-\cos(s\varphi+t\alpha+\delta_l-\delta_{l'})] + (sd+b_l-b_{l'})^2}\right)\right.$$
$$\left. - \tilde{G}\left(\gamma_2,\sqrt{2R^2[1-\cos(s\varphi+t\alpha+\delta_l-\delta_{l'})] + (sd+b_l-b_{l'})^2}\right)\right]e^{ip_2 s}e^{i\alpha\nu_2 t}.$$

In comparison with the case of the right polarization, the selection rules change to $\nu_2 = \nu_1 - 1$, $p_2 = p_1 - \varphi$. As above, we obtain

$$\alpha^L(\omega) = \frac{8\pi}{c\hbar n}\frac{L}{T}\sum_{\nu=0}^{M-1}\frac{|\tilde{g}^L(p,\nu,p-\varphi,\nu-1)|^2}{\left|\frac{\partial}{\partial p}(E_2(p-\varphi,\nu-1) - E_1(p,\nu))\right|}\Bigg|_{p=p(\omega,\nu)},$$

where $p(\omega,\nu)$ is found from the condition $E_2(p-\varphi,\nu-1) - E_1(p,\nu) = \hbar\omega$.

### 4.4. An example

As an example, we consider a tube (15,14). The zone scheme (lines 1 and 2), the squared module of the matrix element of electron - photon interaction (line 3), absorption coefficient (dotted line 4) as functions of $p$ are depicted in figure 3 for various polarizations of the incident light. The same data for the parallel polarization are shown in figure 3 (a). The lines in figure 3 (b) are related to the case of the right-polarized light propagating along the tube axis. Here, the upper band 2 is shifted with respect to the lower band 1 by $+\varphi$, in accordance with the selection rules.

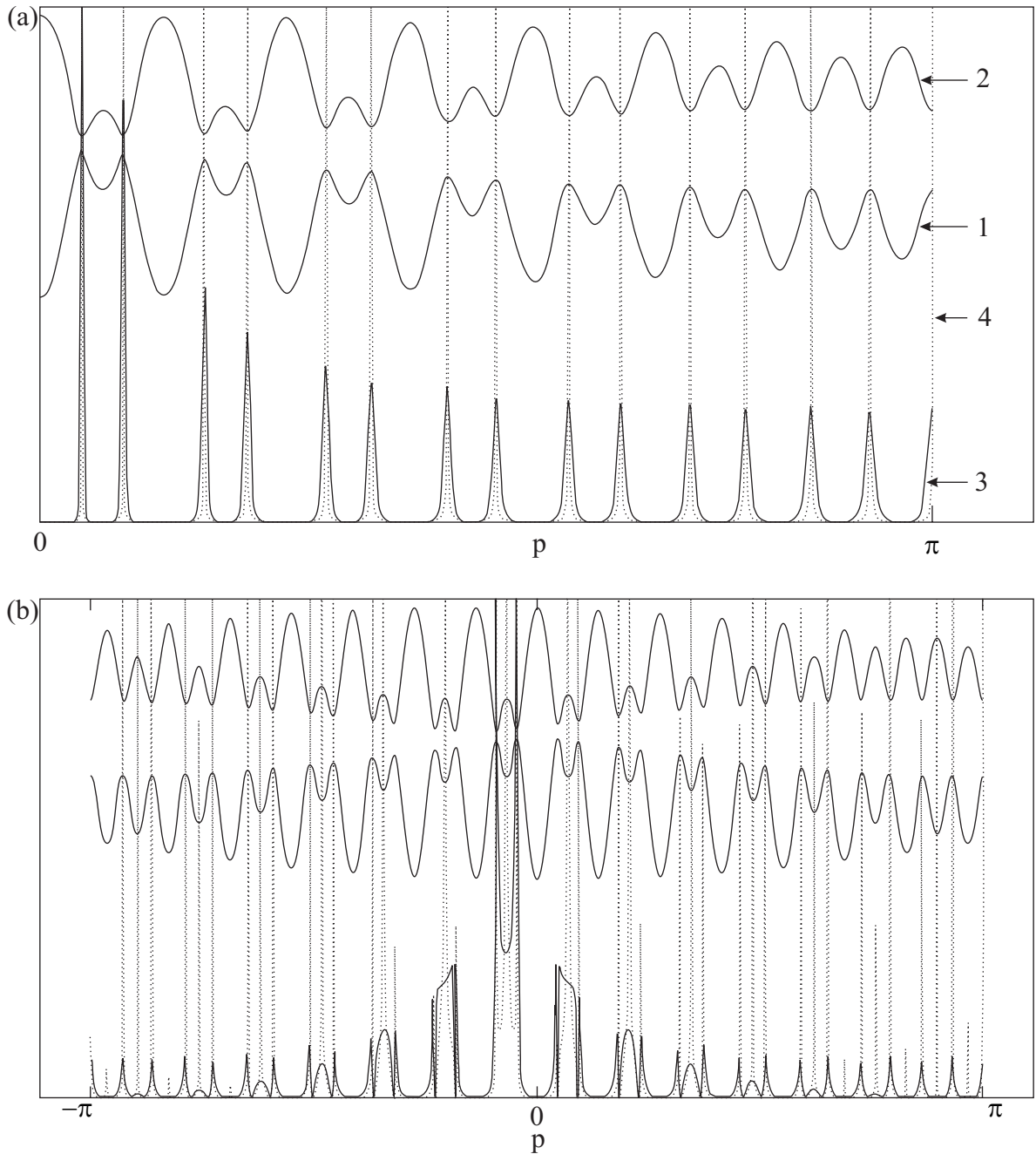

**Figure 3.** The band and absorption spectra for a nanotube (15,14) : (a) parallel polarization, (b) the right circular polarization.

The lines for the left circular polarization are mirror reflections of those for the right polarization with respect to zero. Indeed, let's replace $p$ with $-p$ everywhere in the expression for the absorption coefficient for the right polarization. Then the matrix element of electron - photon interaction $\tilde{g}^R(p, p+\varphi)$ becomes $\overline{\tilde{g}^L(p, p-\varphi)}$, the joint density of states $\left|\frac{\partial}{\partial p}(E_2(p+\varphi) - E_1(p))\right|^{-1}$ becomes $\left|\frac{\partial}{\partial p}(E_2(p-\varphi) - E_1(p))\right|^{-1}$, and the condition

$E_2(p+\varphi) - E_1(p) = \hbar\omega$ becomes $E_2(p-\varphi) - E_1(p) = \hbar\omega$. Note that here the $\nu$-dependence drops out since $M=1$ for the tube (14,15). In general, by replacing $p$ with $-p$ and $\nu$ with $-\nu$ the matrix element of electron - photon interaction $\tilde{g}^R(p, p+\varphi, \nu+1)$, the joint density of states $\left|\frac{\partial}{\partial p}(E_2(p+\varphi, \nu+1) - E_1(p,\nu))\right|^{-1}$ and the condition $E_2(p+\varphi, \nu+1) - E_1(p,\nu) = \hbar\omega$ become $\overline{\tilde{g}^L(p, p-\varphi, \nu-1)}$, $\left|\frac{\partial}{\partial p}(E_2(p-\varphi, \nu-1) - E_1(p,\nu))\right|^{-1}$ and $E_2(p-\varphi, \nu-1) - E_1(p,\nu) = \hbar\omega$, respectively. One can see that the absorption coefficients for the right- and left-polarized light are equal; hence, the circular dichroism is absent within the framework of the considered approximation. Note that this conclusion is not obvious for chiral tubes. In [8], calculations of the optical properties of nanotube (4,2) revealed no circular dichroism either. However, a very slight circular dichroism was found in theoretical work [9]. Samsonidze *et al* [10] suppose that the circular dichroism occurs only when the time-reversal symmetry is broken because of the spatial inhomogeneity of the optical field and an axial magnetic field.

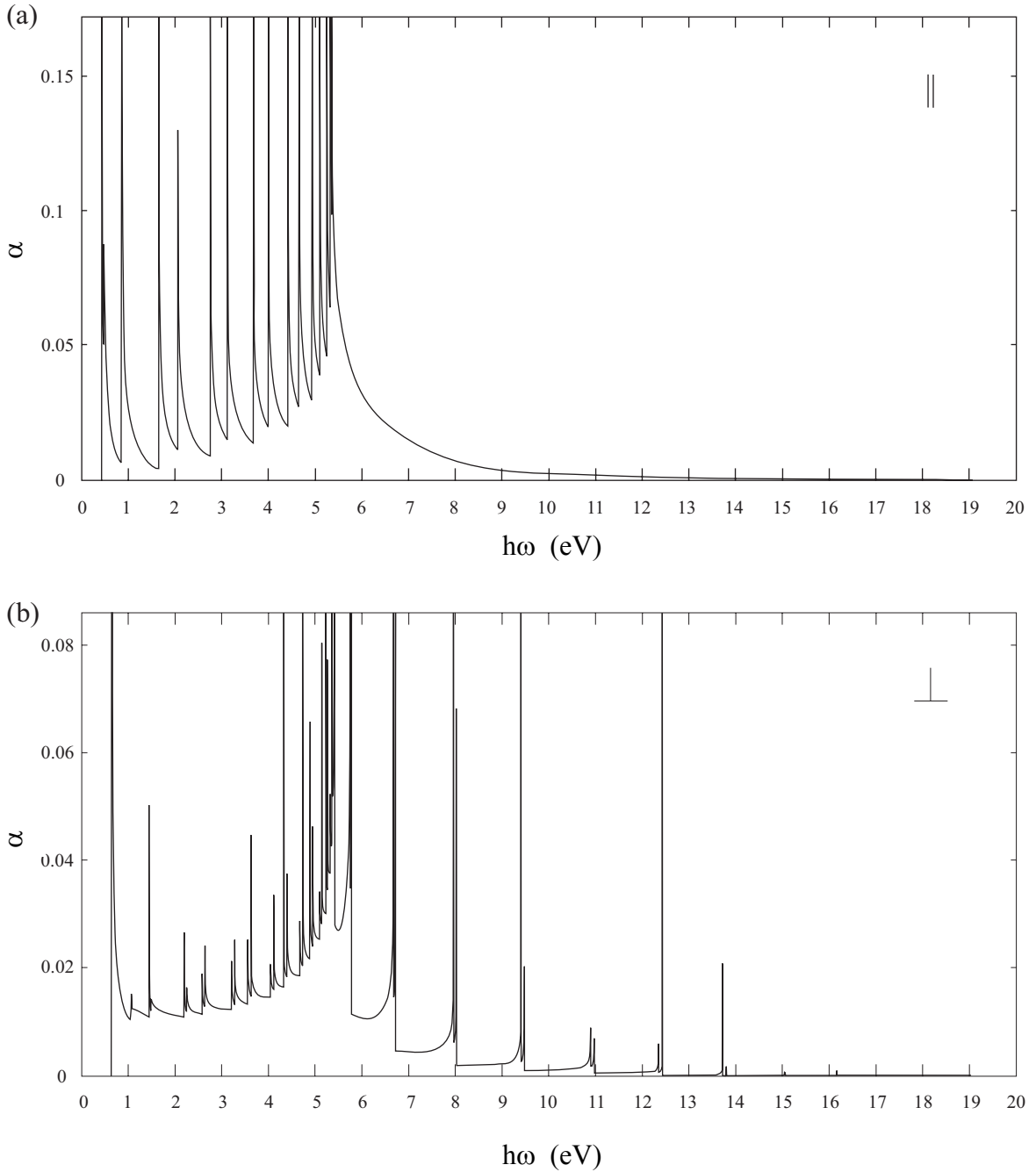

**Figure 4**. Dependence of the absorption coefficient of the isolated nanotube (15,14) on the energy of the incident light: (a) copolarized, (b) cross-polarized.

Let's consider the light linearly polarized, for certainty, parallel to the $x$ - axis. Then

$$|g_x(p_1,\nu_1,p_2,\nu_2)|^2 = \frac{1}{4}|g^R(p_1,\nu_1,p_2,\nu_2) + g^L(p_1,\nu_1,p_2,\nu_2)|^2$$

$$= \frac{1}{4}\Big(|g^R(p_1,\nu_1,p_2,\nu_2)|^2 + |g^L(p_1,\nu_1,p_2,\nu_2)|^2$$

$$+ g^R(p_1,\nu_1,p_2,\nu_2)\overline{g^L(p_1,\nu_1,p_2,\nu_2)} + \overline{g^R(p_1,\nu_1,p_2,\nu_2)}g^L(p_1,\nu_1,p_2,\nu_2)\Big) .$$

Taking into account that $(g^R(p_1,\nu_1,p_2,\nu_2) = \tilde{g}^R(p_1,\nu_1,p_1+\varphi,\nu_1+1)\tilde{\delta}_{\nu_1+1,\nu_2}\delta_{p_1+\varphi,p_2})$ and $(g^L(p_1,\nu_1,p_2,\nu_2) = \tilde{g}^L(p_1,\nu_1,p_1-\varphi,\nu_1-1)\tilde{\delta}_{\nu_1-1,\nu_2}\delta_{p_1-\varphi,p_2})$, the third and fourth terms vanish. Evidently, $|g_x(p_1,\nu_1,p_2,\nu_2)|^2 = |g_y(p_1,\nu_1,p_2,\nu_2)|^2$. Then the absorption coefficient of linear cross-polarized light is

$$\alpha_\perp(\omega) = \frac{1}{4}\big(\alpha^R(\omega) + \alpha^L(\omega)\big) = \frac{1}{2}\alpha^R(\omega).$$

Notice that the selection rules for the light linearly polarized parallel and perpendicular to the nanotube axis are different. The differences manifest themselves in the light absorption (linear dichroism) (figure 4). A distinctive feature of the absorption of the cross-polarized light is that the transition at the frequency corresponding to the band gap is forbidden. Therefore, the absorption edge in this case is shifted to higher frequencies, which is clearly seen in figure 4.

## 5. Discussion and comparison with the absorption by a graphite plane

The absorption spectrum of polarized light revealed by a system of single-wall nanotubes, vertically aligned on a quartz substrate, was experimentally investigated in [5] within the frequency range 0.5-6 eV. The average diameter of the nanotubes was $\sim 2.0$nm, with the standard deviation of $\sim 0.4$nm. Evidently, for nanotubes with various diameters and chiral angles there is a noticeable scatter in the peak positions on the absorption lines, which actually causes the observable line to smooth out. Moreover, the nanotubes investigated in [5] were not ideally rectilinear. Because of these, sharp peaks are difficult to see on the absorption lines obtained in [5]. However, such peaks were observed in other experiments [11-14], in which the absorption lines were measured for energy up to 3 eV. As already mentioned, the wider frequency range was considered in [5], and the distinct absorption maxima were found to occur at 5.25 eV for the cross-polarized and at 4.5 eV for the copolarized light. The nature of these maxima is not quite clear. Some authors explain them by excitation of π–plasmons [15,16]. If we smooth out our absorption lines (figure 4), the absolute absorption maximum appears at frequency close to $5.25 eV$, and, in accord with [5], the absorption intensity of the copolarized light proves to be greater than that of the cross-polarized one. However, the observed maximum at 4.5 eV on our smoothed curve does not appear.

To understand the structure of the absorption lines, we have studied, by using the ZRP method, the optical absorption spectrum for the infinite graphite plane. On the basis of the ZRP method, it is possible to find the dispersion equation and thus to determine the band structure of the π-electrons of the infinite graphite plane (figure 5). A similar band structure was obtained by tight-binding method [17-22].

As the unit cell of a hexagonal lattice of a two-dimensional (2D) graphite plane contains two atoms, the band structure contains two bands. The Brillouin zone is a hexagon. The point Γ corresponds to its centre. The energy has a maximum (minimum) at this point. M is the saddle point corresponding to the midpoints of hexagonal edges. The points K, where the bands touch, are at the vertices of the hexagon, that is, the 2D graphite is a semimetal. Following [17,19,20,22], the band structure of an (*n,m*) nanotube can be obtained by imposing periodic boundary conditions on the orthogonal component of the wave vector with respect to the nanotube axis; i.e. the wave vector should satisfy the condition

$$\mathbf{Ck} = 2\pi f,$$

where *f* is an integer. Hence the allowed values of the wave vector lay on *n+m* lines inside the Brillouin zone of the graphite (figure 6). Note that due to a large number of atoms in the nanotube unit cell, the first Brillouin zone of the tube is a small segment containing the point Γ.

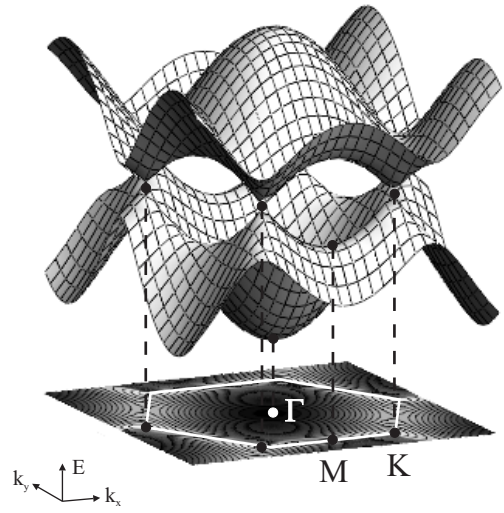

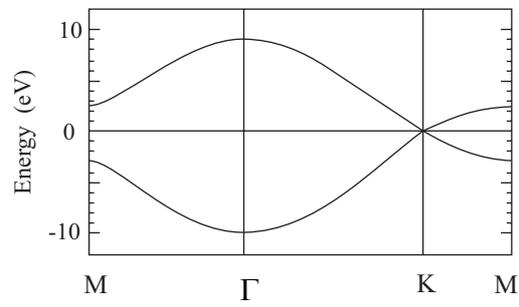

**Figure 5**. The band structure of an infinite graphite plane.

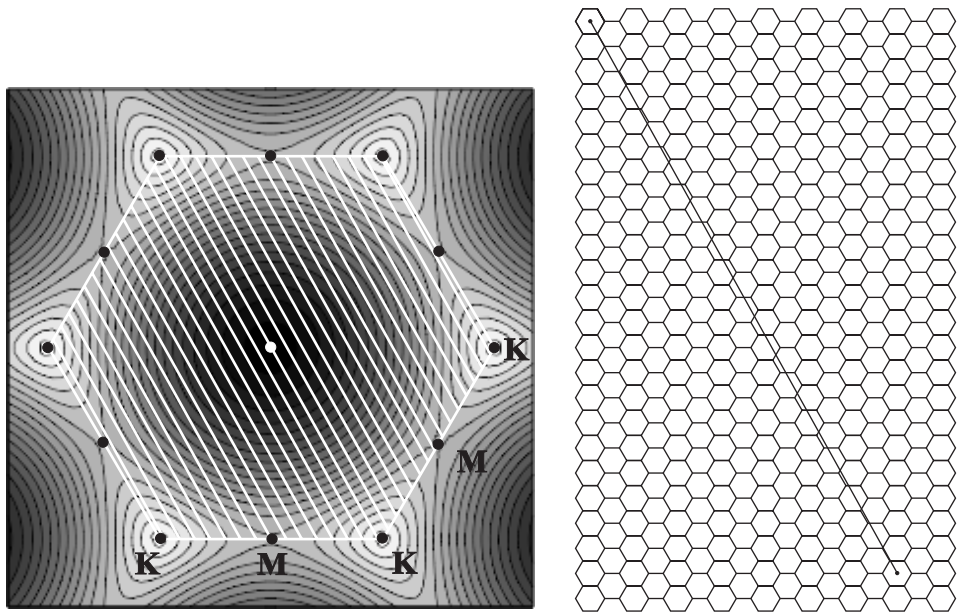

**Figure 6.** Allowed values of the wave vector for a nanotube (15,14).

Extending a line of allowed values for *k* from Γ to the centre of another Brillouin zone of graphite, we get an extended zone scheme. On its path through other Brillouin zones to point Γ of another Brillouin zone, this line regenerates the allowed segments of the 1st zone with different *f*. If on this path all the allowed segments of the first zone are regenerated, such a nanotube is described by two spirals (figure 6) and their extended zone scheme contains only two bands (as, for example, for a nanotube (15,14), figure 2).

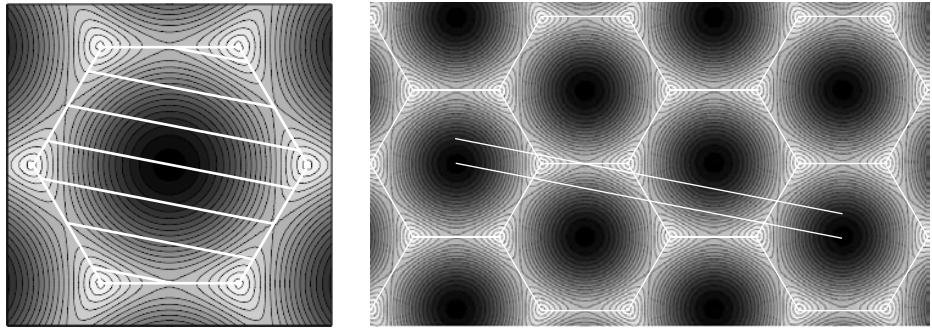

**Figure 7.** Allowed values of the wave vector for a nanotube (4,2).

However, the two-spiral structure is not inherent to all nanotube. For example, for a nanotube (4,2) two lines with $f=0$ and $f=1$ are necessary to get, as above, all the allowed segments of the Brillouin zone; as a result, the band structure contains 4 bands (figure 7) and the spatial structure of this tube is described by four spirals.

Let's return to a nanotube (15,14) and figure 6. Two bands correspond to one line of allowed values of the wave vector *k*. The maxima of the energy of the bottom band are located at the intersections of the line of the allowed values of *k* with the edges of the hexagons, and the minima are in between them inside each crossed hexagon. Therefore, the nanotube bottom band is a sequence of alternating maxima and minima (figure 2). The closer that line passes to the centre of a hexagon (point Γ), the deeper the corresponding minimum is. The top band is just a mirror copy of the bottom one, the minima of the former being located exactly opposite to the maxima of the latter and vice versa. The greatest distance between the maxima of the bottom band and the minima of the top band are at the intersection of the line of the wave vector allowed values with the midpoints of the edges (point M) and corresponds to 5.32 eV. If the line passes near this point, so that a slight shift occurs, then, accordingly, this distance appears to be somewhat lesser. When moving from the midpoint of a hexagon edge to a hexagon vertex, this distance decreases and the bands touch at K. However, in a real nanotube, because of its curvature, the bands are a little distorted and a narrow gap of ~0.01eV (depending on the nanotube diameter) appears. This is well seen from the zone scheme of a nanotube (15,14) (figure 2), where the envelopes of the bottom band maxima and the top band minima coincide with the energy lines in the zone scheme of graphite while the wave vector runs over the segment KM on the hexagon edge (figure 5). As the extremum points of both bands of the nanotube occur at identical values of the wave vector, the electronic joint density of states becomes infinite at these points. This results in the Van Hove singularities (peaks) on the absorption lines, the peaks corresponding to transitions between the extremum points. Some of these transitions are forbidden because of zero value of the matrix element of electron - photon interaction at these points.

Notice that the above reasoning is used here only for interpretation of the results obtained on the basis of direct calculations with the use of the ZRP method and, therefore, with a complete account of the spatial structure of nanotubes.

In the case of the parallel polarization, the peaks are wider and more distinct than those in the case of the cross polarization. As the frequency of incident light increases, the density of peaks increases, reaches a maximum around 5.3 eV, and then drastically falls. The result is that the maximum of the light absorption is somewhere in this region. This happens because, due to a decrease in the slope of the curve KM, the density of extremum points of both bands increases as the centre of an edge (point M) is approached.

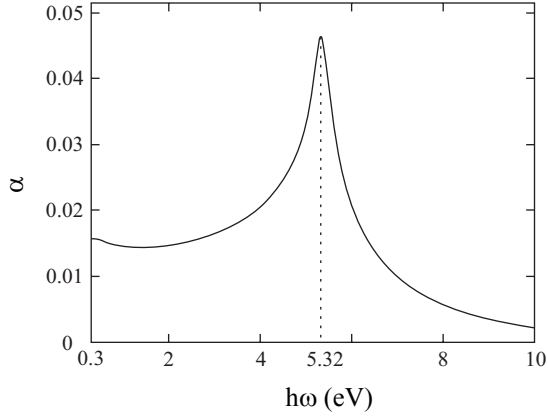

**Figure 8.**
Absorption spectrum of an infinite graphite plane for direct interband transitions of π-electrons, obtained on the basis of the ZRP method.

The frequency dependence of light absorption caused by direct interband transitions of π-electrons for the graphite plane is shown in figure 8. The absorption maximum is close to the energy that equals the distance between the bands at point M. There is some discrepancy between the light absorption spectra of graphite calculated by the ZRP method and obtained experimentally in [23-25]. Namely, the ZRP line has a peak at 5.32 eV, while the analogous experimental line has this peak shifted to 4.6 eV. The value of 4.6 eV is attributed to the distance between the π−bands at the saddle point M [24]. The same distance obtained here by the ZRP method is 5.32 eV. The cause of this discrepancy may be related to the choice of the universal fitting parameter $\mu$. It was chosen from the requirement that the distance between the two nearest Van Hove singularities on both sides from the edges of the narrow forbidden band in a nanotube zigzag (15,0), which was calculated by the ZRP method, and that obtained experimentally in [4] be equal. The use of this fitting allowed us to achieve rather good results near the forbidden bands for different nanotubes, that is, near the K points of the graphite bands. However, away from this region, certain discrepancies in energies of the bottom and top bands may occur. These discrepancies may add together to total as high as 0.72 eV at the saddle point *M*. Actually, it is difficult to expect a good quantitative agreement within the whole energy range for a model with one fitting parameter. A similar situation is observed for the tight-binding model with a single parameter (overlap energy) $\gamma_0$ [26], where the distance between the bands at the saddle point is $2\gamma_0$, $\gamma_0 \in (2.5-2.9) eV$. To improve the tight-binding model, the three-nearest neighbour approximation with additional parameters was considered [26]. Note that the ZRP method with one parameter, in general, takes into account influence of all neighbours. Particularly, it results in the appearance of a suitably narrow forbidden band for some "metal" nanotubes [3] while within the tight-binding approximation such a band appears for the additional accounting of σ−π hybridization.

By varying the fitting parameter in the ZRP method and requiring that the interband distance at the saddle point M for the graphite plane and that experimentally obtained be equal, we can achieve a good coincidence of our calculated absorption line and that obtained experimentally in the vicinity of 4.6 eV. Similar results are also obtained there for the case of the absorption copolarized light by nanotubes. But then the discrepancy near the point K with the scanning tunnelling spectroscopy data appear. The coincidence of the location of the absolute maximum of absorption for the graphite plane and that for nanotubes seems clear for the case of the copolarized light. Generally, the corresponding absorption lines for the tubes, especially for tubes of sufficient diameters, should not differ much from those for the graphite plane, except for the presence of singularities due to the one-dimensional nature of nanotubes. Therefore, the maximum of the absorption at 4.6 eV [5] is simply a property of graphite. The absorption of the cross-polarized light by carbon tubes is not related directly to the optical properties of graphite, for the graphite plane does not absorb cross-polarized light.

Note that a maximum at 5.25eV for the absorption of the cross-polarized light is also observed; each peak of the corresponding calculated line is slightly blue-shifted as compared to the case of the copolarized light absorption, though not so much as to provide a shift of 0.65 eV from the characteristic graphite maximum at 4.6 eV. Therefore, the nature of this maximum is not clear yet.

Also, notice that the ZRP results on the co- and cross-polarized absorption spectra for the nanotubes (6,5), (7,5), (8,4), (7,6) are in a quite satisfactory agreement with the data obtained in [27] by the polarized photoluminescence excitation spectroscopy.

It is necessary to mention that in this work only the subsystem of tube π-electrons was taken into account. It was shown in [23] that the σ-electrons do not contribute to the light absorption in graphite in the energy region below ~9eV. However, it was stated in [24] that the σ-electrons of graphite may affect the absorption spectrum already near ~6eV. As it was emphasized in [28], in small nanotubes such as (5,0), (2,4) and (3,3), thoroughly studied there, a large curvature effect may also lead to the hybridization of the π and σ orbitals. With account of the hybridization effect the tube (5,0), according to calculations [28] and [29], appears to be metallic. Our calculations of the optical absorption spectra for tubes (5,0), (4,2) with account only the π−electrons reveal sharp peaks at 2.18 eV, 2 eV and 3.6 eV, 3.52 eV respectively, which are not in bad agreement with experimental values 2.1 eV and 3.1 eV obtained in [28] for arrays of these nanotubes aligned in channels of an $AlPO_4$-5 single crystal. However, we did not find the sharp peak at 1.37 eV obtained experimentally and any peaks at all in the absorption spectra of a tube (3,3) in the photon energy range 0.5-4 eV. We can not exclude that some peaks at the low edge of the absorption line may appear because of exciton generation.

In principle, the ZRP method also allows one to incorporate the σ - electrons and, by increasing the number of fitting parameters, to improve quantitative coincidence with experimental data.

## 6. Conclusion

The version of the method of zero-range potentials with one universal fitting parameter proposed in the present work proves to be an effective tool for qualitative and, at least, near the forbidden bands, quantitative description of the electronic properties of all single-wall carbon nanotubes and the graphite plane. Being, in fact, an analogue of the Kronig-Penny model for nanotubes, it permits one to grasp in rather simple way all the effects and contributions mostly originated from the geometrical structure of nanotubes. In combination with the model of a tube as a system of parallel monoatomic spirals, the zero-range potential method works well when the explicit dispersion equations and wave functions of the band π-electrons for a single-wall nanotube and the contributions to the absorption spectra of polarized light caused by direct interband transitions of the π-electrons in isolated single-wall nanotubes are to be obtained.

In the framework of this approximation, the circular dichroism for the light propagating along the ideal chiral nanotube does not appear. In the case of the copolarized and cross-polarized light falling perpendicular to the tube axis, the linear dichroism should be observed, due to an essential distinction in the selection rules. For the cross-polarized light, it may manifest itself as a blueshift of the absorption edge; for the copolarized light, the peaks of the absorption line are wider and more intensive. These conclusions agree qualitatively with experimental results [5]. The smoothed absorption lines for the parallel polarization qualitatively coincide with the absorption lines for the graphite plane, but the characteristic maximum shifted from 4.6 eV [24-26] to 5.3 eV. This distinction can be caused by both limitations of the one-parameter model and ignoring the effects of σ−π hybridization for higher energies.

## Acknowledgments
The authors are grateful to M. Sushko for valuable discussions and remarks.